# Generation of Spectrally Pure Microwave Signals

*Eugene N. Ivanov and Michael E. Tobar*

*ARC Centre of Excellence For Engineered Quantum Systems, Department of Physics, University of Western Australia, 35 Stirling Highway, Crawley WA 6009, Australia.*

**Abstract.** Based on the principles of microwave circuit interferometry we've constructed a Reduced Noise Amplifier (RNA) with power spectral density of phase, as well as amplitude, fluctuations close to -170 dBc/Hz at 1 kHz offset. The RNA has been incorporated with a cryogenic sapphire resonator as a loop oscillator whose noise performance is governed by the Leeson's model. Following this model and using the results of *in-situ* measurements of gain and phase fluctuations of the RNA we inferred noise properties of the oscillator. In particular, for a signal transmitted through the resonator we found that power spectral density of its phase/ amplitude fluctuations must be close to -185 dBc/Hz at offsets above 300 Hz. We discuss a few approaches that will allow direct measurements of such low levels of noise.

1. INTRODUCTION

Microwave signals with the best short-term frequency stability and low phase noise are currently derived from the optical sources [1-3]. This is the result of the outstanding progress in laser frequency stabilisation: lasers with fractional frequency instability of 5 10$^{-17}$ at 1 s operate in a number of research labs [4-6]. In parallel with the development of ultra-stable lasers, a technique of optical frequency division was also perfected [7-11]. The fractional frequency error associated with the optical frequency division is small enough to preserve the long-term frequency stability of any state-of-the-art optical clock when transferring its signal to the microwave domain. Furthermore, with a proper choice of laser power, detector load impedance and bias voltage, it is possible to generate X-band signals with the SSB phase noise close to -160 dBc/Hz at 1 kHz offset [1]. This is approximately 3 dB better than what was achieved with the high-power Sapphire Loaded Cavity (SLC) oscillators [12].

The Opto-Electronic-Oscillator (OEO) is another example of using lasers for generation of low-phase noise microwaves [13-15]. The OEOs, in terms of their phase noise performance, are inferior to both SLC oscillators and the optical frequency dividers. Among their advantages are relatively low complexity, low cost and compactness. In the past few years, serious advances were achieved in the manufacturing of optical micro-resonators, as well as in the understanding of their intrinsic noise processes.

The phase noise performance of the optical frequency dividers have also been improving steadily since the key noise mechanisms of the optical-to-microwave conversion were understood [8]. On the other hand, the progress in the field of sapphire crystal oscillators was rather slow. The situation changed when it was realised that cryogenically cooled sapphire crystals can be used for noise filtering. Consider, for example, a loop oscillator based on conventional microwave amplifier with the SSB phase noise -100 $\mathcal{F}^{-1}$dBc/Hz, where $\mathcal{F}$ is the offset (or Fourier) frequency. Assuming that the frequency selective element of the loop oscillator is a 10 GHz cryogenic sapphire resonator with the loaded Q-factor of 1 billion, the SSB phase noise of the microwave signal transmitted through the resonator, according to Leeson's model [16], must be close to -180 dBc/Hz at Fourier frequencies $\mathcal{F}$ > 3 kHz. This noise level corresponds to the Standard Thermal Noise Limit for the microwave signal with power of 1 mW [17].

As follows from the above example, cryogenic sapphire oscillators can be considered as a viable alternative to the photonic microwave sources. Indeed, they offer (i) lower noise, (ii) cleaner noise spectra due to low vibration sensitivity and (iii) long maintenance-free operation limited by the lifetime of the cryocooler [18]. In this work we consider noise properties of one practical implementation of the cryogenic sapphire oscillator, which we refer to as the Reduced Noise Oscillator (RNO).

2. REDUCED NOISE OSCILLATOR

Schematic diagram of the RNO is shown in Fig. 1. It consists of a cryogenic sapphire resonator and Reduced Noise Amplifier (RNA).

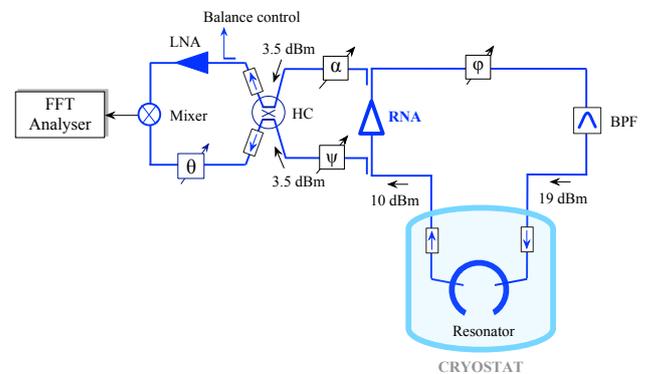

Fig. 1. Schematic diagram of Reduced Noise Oscillator: RNA- Reduced Noise Amplifier, BPF – band-pass filter, LNA – low-noise amplifier

An external readout system is also shown, which allows *in-situ* measurements of either phase or amplitude fluctuations of the RNA depending on the value of the phase delay, θ.

### 2.1. Sapphire Resonator

The cryogenic sapphire resonator operates at frequency $f_{res}$ ~ 11.343 GHz. Its temperature was kept close to the frequency-temperature turning point $T_{TP}$ = 6.8 K in order to improve consistency of noise measurements at low Fourier frequencies [19]. The coupling to the resonator was set in such a way as to minimise its transmission loss: the total power loss between the ports of the cryostat was approximately 9 dB.

The resonator was heated up to its operating temperature mostly by the dissipated microwave power (~ 20 mW). Only an additional 4 mW was needed to be supplied by the heater of the temperature control system.

Since the resonator was strongly coupled there was a relatively large discrepancy between the intrinsic $Q_0$ and loaded $Q_L$ quality factors. For the resonator in question, we had $Q_0$ ~ 790 million and $Q_L$ ~ 195 million. For a given value of the $Q_L$, the resonator loaded half-bandwidth is $\Delta f_{0.5} = f_{res}/(2Q_L)$ ~30 Hz. This corresponds to the characteristic frequency beyond which technical fluctuations of the transmitted signal are low-pass filtered at the rate of 6 dB/Octave.

### 2.2. Reduced Noise Amplifier

The schematic diagram of Reduced Noise Amplifier (RNA) is shown in Fig. 2.

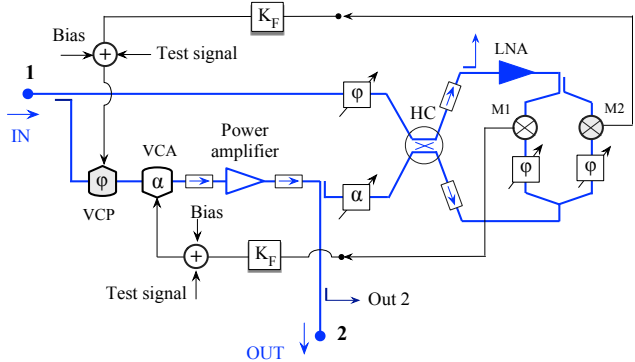

Fig. 2. Reduced Noise Amplifier: HC -3dB Hybrid Coupler

It features microwave Mach-Zehnder interferometer. One arm of the interferometer contains a high gain power amplifier, a Voltage Controlled Attenuator (VCA) and a Voltage Controlled Phase-shifter (VCP) to allow electronic feedback control. Another arm of interferometer contains a variable mechanical phase-shifter. The interferometer is followed by a dual-channel readout system, one channel of which is tuned to be sensitive to phase fluctuations inside interferometer, while another channel is optimized to detect amplitude fluctuations. Two feedback control loops keep the interferometer balanced by continuously adjusting parameters of the VCA and VCP. This ensures small-signal operation of the Low Noise Amplifier (LNA) at the interferometer "dark port" and, as a consequence - the highest spectral resolution of the dual-channel readout. In such a regime, the microwave signal transmitted through the RNA sees an amplifier with an effective noise temperature close to that of the LNA.

To study the RNA noise properties, the interferometric setup shown in Fig. 3 was implemented. The microwave interferometer was balanced manually by varying insertion loss α and phase delay ψ in its arms. Depending on the phase difference between two mixer ports (adjusted with the phase delay θ in Fig. 3) the voltage noise at its output varied synchronously either with phase or amplitude mismatch fluctuations of the interferometer. In each case, the voltage noise spectrum at the mixer output was measured. Multiple measurements were performed especially at low Fourier frequencies (0<ℱ<1 Hz) to ensure that noise spectra were free from the environmental disturbances.

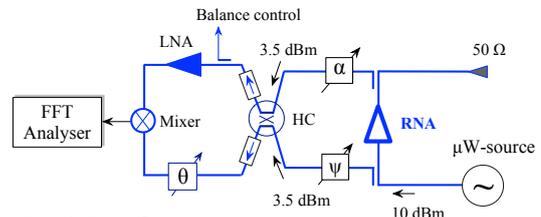

Fig. 3. Interferometric noise measurements system

In order to convert measured voltage noise into either phase or amplitude fluctuations of the RNA, the readout system was calibrated. Its phase sensitivity was measured to be approximately 7 V/rad. The amplitude sensitivity was taken to be 7 V, as for the interferometric systems, both sensitivities are equivalent [20].

The noise spectra of the RNA, as well as the noise floor of the readout system (measured with 50Ω termination at the LNA input), are shown in Fig. 4.

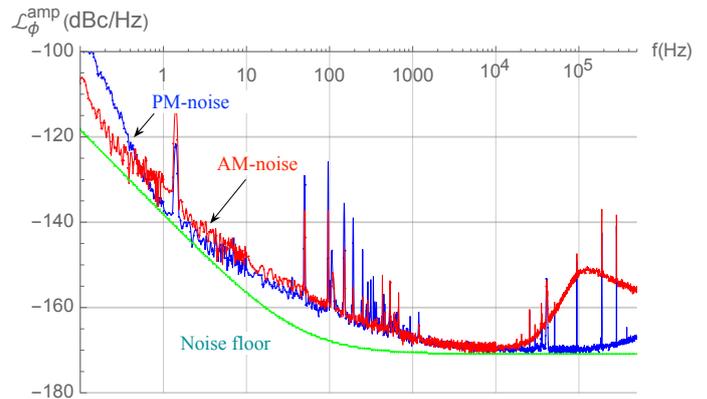

Fig. 4. RNA measured noise spectra

Excess AM-noise was observed at Fourier frequencies above 30 kHz, due to the relatively slow time response of the VCA (its bandwidth was measured to be ~ 300 kHz). For comparison, the bandwidth of the phase control loop was approximately 3 MHz. This is because time response of the VCPs is much faster that of the VCAs.

The aforementioned noise measurements were also performed *in-situ* with the RNA acting as a gain stage of the loop oscillator (see Fig. 1). No noticeable discrepancies with the data in Fig. 4 were observed.

## 2.3. Reduced Noise Oscillator

Following in Leeson's "footsteps" [16] we derived the following relations between phase/amplitude fluctuations of the loop amplifier (which is the RNA, in our case) and phase/amplitude fluctuations of the microwave signal transmitted through the resonator:

$$S_\phi^{transm}(\mathcal{F}) \simeq S_\phi^{amp}(\mathcal{F}) \left(\frac{\Delta f_{0.5}}{\mathcal{F}}\right)^2 + \frac{k_B T_o}{2\, P_{transm}} \quad (1)$$

$$S_\alpha^{transm}(\mathcal{F}) \simeq S_\alpha^{amp}(\mathcal{F}) \frac{1}{\chi^2 + (\mathcal{F}/\Delta f_{0.5})^2} + \frac{k_B T_o}{2\, P_{transm}} \quad (2)$$

Parameter $\chi$ in Eq. (2) characterizes saturation of the loop amplifier: $\chi = \frac{U}{K_{amp}} \frac{dK_{amp}}{dU}$, where $K_{amp}$ is the gain of the amplifier and $U$ is the stationary amplitude of the input signal. For the loop amplifier in saturation $\chi \approx -1$. The last term in Eqns. (1, 2) is the thermal noise limit which depends on the Boltzmann constant $k_B$, ambient temperature $T_o$ and signal power $P_{transm}$.

Fig. 5 shows oscillator noise spectra inferred from those of the RNA in accordance with Eqns. (1, 2). Both phase and amplitude noise spectra converge to -185 dBc/Hz at $\mathcal{F} > 300$ Hz. This is the thermal noise limit at signal power of 3 dBm (see Fig. 1). Apart from the bright spectral line at 1.5 Hz due to the cryocooler operation, the RNO noise spectra are much cleaner than those of the photonic oscillators in [1, 2].

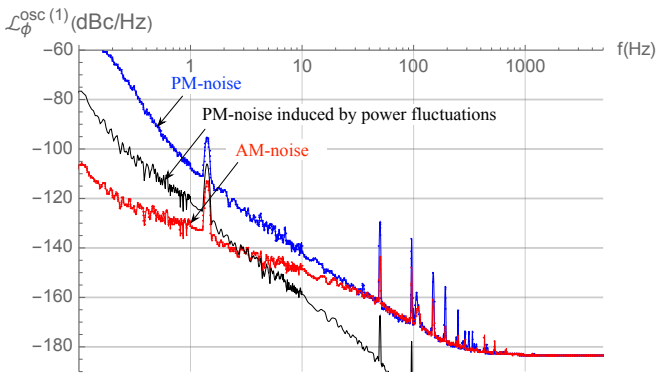

Fig. 5. Inferred noise spectra of transmitted signal

Eq. (1) was derived assuming that resonant frequency $f_{res}$ does not fluctuate. In fact, the $f_{res}$ is a subject to fluctuations of both temperature and microwave power [21]. As for the ambient temperature fluctuations, they are too slow and weak (due to thermal shielding) to have a noticeable effect on oscillator phase noise at Fourier frequencies $\mathcal{F} > 1$ Hz. As for the power fluctuations, their effect on oscillator phase noise is shown in Fig. 5. It was computed based on the given resonator sensitivity to dissipated microwave power: $df_{res}/dP_{diss} \sim 0.1$ Hz/mW.

The absence of the direct noise measurements of the RNO can be viewed as a weakness of this work. Below we'll discuss possible approaches to allow direct measurements of the weak phase fluctuations, as well as whether such measurements are necessary in case of the RNO.

First, we consider an interferometric two-oscillator noise measurements system in Fig. 6. It has adequate sensitivity to detect both phase and amplitude fluctuations of the RNO [20]. Yet it must be ruled out, as it is not possible to tune two cryogenic sapphire resonators within a few Hz of each other. In practice, the smallest frequency difference between two nominally identical sapphire resonators we have observed when cooled to 4 K, was close to 1 MHz.

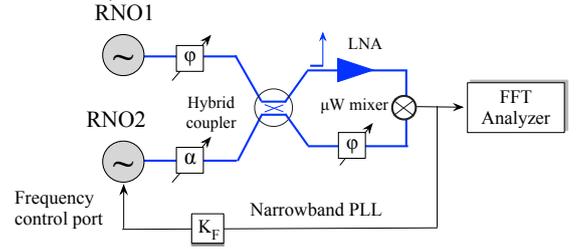

Fig. 6. Interferometric noise measurements system

Secondly, since we operate a pair of ~11 GHz Pound-stabilized Cryogenic Sapphire Oscillators (CSO) [23,24], one may suggest using the cross-correlation noise measurements system in Fig. 7a similar to that employed in [1, 2] for characterization of phase fluctuations of the photonic oscillators. The problem is that, even the use of cross-correlation would not be able to "bridge" more than 25 dB gap between the spectral densities of phase fluctuations of the RNO and CSO existing at Fourier frequencies beyond 10 Hz (see Fig. 8). In other words, the CSOs are too noisy to serve as auxiliary oscillators in the three-oscillator comparison experiment.

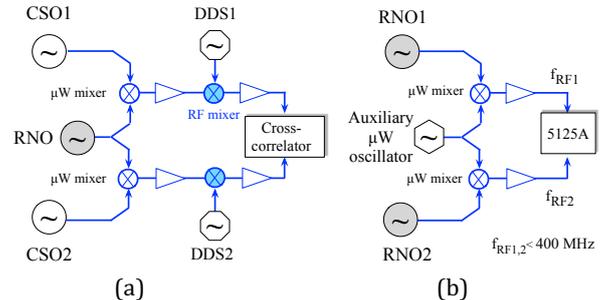

Fig. 7. Noise measurements systems

One may suggest measuring the RNO phase noise with the apparatus in Fig. 7b. It requires the use of two RNOs and an auxiliary oscillator. Provided that frequencies of the two beat notes are less than 400 MHz, a phase noise test set 5125A from Symmetricom can be used to measure spectral density of their phase difference fluctuations [25]. The problem is that the spectral resolution of such noise measurements would be limited by technical fluctuations of the microwave mixers. Again, referring to the noise spectra in Fig. 8, we conclude that the readout system in Fig. 7b is incapable of detecting RNO phase noise at $\mathcal{F} > 10$ Hz.

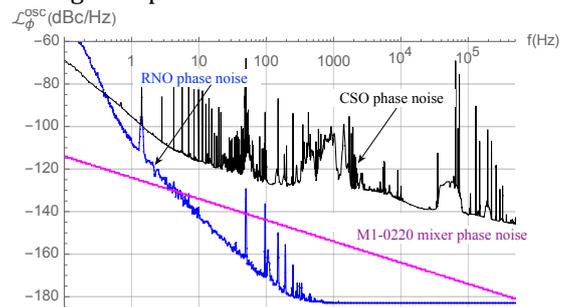

Fig. 8. Phase noise spectra of various oscillators

It appears that the only way to directly measure phase noise of the RNO is via the use of cross-correlation system in Fig. 7a featuring three RNOs of comparable noise performance. This means repeating the massive effort of the "photonic team" in [1, 2] who had to build three photonic oscillators in order to figure out phase noise of one of them. At this stage it makes sense to pause and ask a question: Whether such an effort will give us any new information about the RNO noise properties in addition to what we already know? The answer, we believe, is negative. *This is because, in contrast to a complex photonic microwave source whose noise properties cannot be unambiguously inferred from a series of auxiliary measurements, the RNO is essentially a classical loop oscillator with the perfectly predictable noise performance. All one has to do in order to characterize RNO noise properties is to measure spectra of phase/amplitude fluctuations of the RNA, along with the parameters/ sensitivities of the cryogenic resonator.*

Switching from the Frequency to the Time Domain we decided to measure RNO frequency stability despite the fact that RNO was not meant to be a frequency-stable source. For that, we carried out a simple two-oscillator comparison experiment involving one of our Pound-stabilized CSOs detuned from the RNO by approximately 100 MHz. The RNO was modified to improve stability of its mean frequency. This was done by controlling temperature of the microwave band-pass filter as shown in Fig. 9. The Allan deviations of the fractional frequency fluctuations of the RNO and CSO averaged over time τ are shown in Fig. 10.

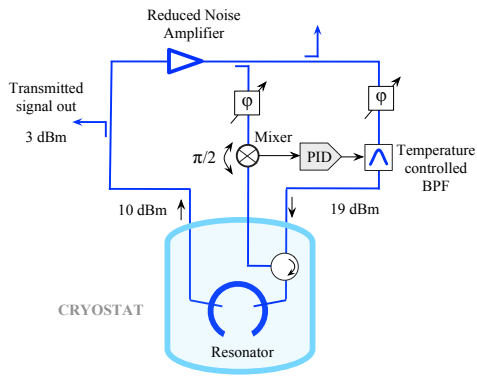

Fig. 9. RNO with mean frequency stabilization

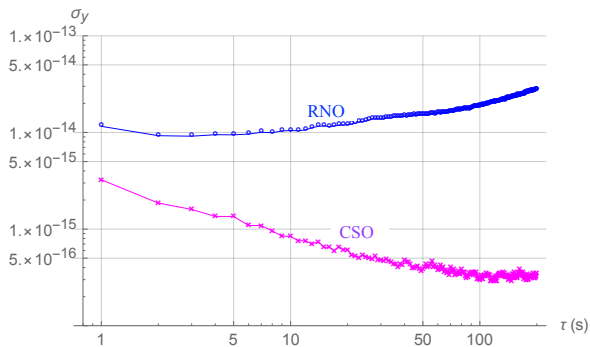

Fig. 10. Fractional frequency stability vs integration time

The results of the above measurements allowed us evaluate the RNO phase noise at low Fourier frequencies.

Using the basic relations from [27] we obtained -57 dBc/Hz for the RNO SSB phase noise at $\mathcal{F}$ = 0.1 Hz. For the CSO the corresponding value was – 70 dBc/Hz. Both values are in reasonable accord with the results of the direct measurements in the Frequency Domain (see Fig. 8).

In principle, the fractional frequency instability of the RNO can be reduced to $3 \cdot 10^{-15}$ for integration times 1s<τ<100s. This estimate comes from our measurements of intrinsic voltage fluctuations of the cryogenic mixer. The latter was composed of two tunnel diode detectors and a 3dB Hybrid coupler. Potentially, one can replace the room temperature mixer in Fig. 9 with the "home-made" cryogenic counterpart in order to minimize the RNO frequency random walk caused by fluctuations of ambient temperature.

3. SUMMARY

First, we reported on the development of the microwave Reduced Noise Amplifier (RNA) with the SSB phase noise close to -140 dBc/Hz at $\mathcal{F}$ =1 Hz and approaching -170 dBc/Hz at $\mathcal{F}$ > 1 kHz. Moreover, at Fourier frequencies of 1 Hz <$\mathcal{F}$< 10 kHz, the spectral density of the RNA amplitude fluctuations was almost indistinguishable from that of its phase fluctuations.

Following this, we assembled the cryogenic sapphire resonator with the loaded Q-factor of 200 million and transmission loss of 3 dB.

Finally, we described the construction of a Reduced Noise Oscillator (RNO) by combining the RNA and the cryogenic sapphire resonator in the oscillations-sustaining loop. From the *in-situ* noise measurements we concluded that the spectral densities of both phase and amplitude fluctuations of the signal transmitted through the resonator approach the thermal noise limit of -185 dBc/Hz at $\mathcal{F}$ > 300 Hz.

Up to now, the progress in oscillator frequency stabilization resulted in a paradoxical situation, where the spectral density of phase fluctuations was reduced well below that of the amplitude fluctuations [12]. This is highlighted by the fact that, there exist no useful techniques that enable efficient cancellation of oscillator amplitude fluctuations. The approach implemented in the RNO design resolves the above paradox by suppressing amplitude fluctuations at their very origin, i.e. in the gain stage of the loop oscillator.

ACKNOWLEDGEMENTS

This work was funded the Australian Research Council grant numbers DP190100071 and CE170100009.